\documentclass{ws-procs975x65}

\begin{document}

\title{VORTICITY FROM ISOCURVATURE IN THE EARLY UNIVERSE}

\author{ADAM J. CHRISTOPHERSON}
\address{School of Physics and Astronomy, University of Nottingham, \\
University Park, Nottingham, NG7 2RD, United Kingdom}

\author{KARIM A. MALIK}
\address{Astronomy Unit, School of Physics and Astronomy, Queen Mary University of London, \\
Mile End Road, London, E1 4NS, United Kingdom}

\begin{abstract}
Vorticity is ubiquitous in nature however, to date, studies of vorticity in cosmology and the early universe
have been quite rare. In this paper, based on a talk in session CM1 of the 13th Marcel Grossmann Meeting,
we consider vorticity generation from scalar cosmological perturbations
of a perfect fluid system. We show that, at second order in perturbation theory, vorticity is sourced by a 
coupling between energy density and entropy gradients, thus extending a well-known feature of classical 
fluid dynamics to a relativistic cosmological framework. This induced vorticity, sourced by isocurvature
perturbations, may prove useful in the future as an additional discriminator between inflationary models.
\end{abstract}


\bodymatter

\section{Introduction}

While in classical fluid dynamics vorticity is a well-studied phenomenon, there have been very few 
corresponding studies in early universe cosmology. In the classical case, vorticity is simply
defined as $\vec{\omega}=\vec{\nabla}\times\vec{v}$, where $\vec{v}$ is the velocity of the fluid,
and evolves according to
\begin{equation}
\frac{\partial\vec{\omega}}{\partial t}=\vec{\nabla}\times(\vec{v}\times\vec{\omega})
+\frac{1}{\rho^2}\vec{\nabla}\rho\times\vec{\nabla}P\,.
\end{equation}
We can see that the rightmost term, the baroclinic term, acts as a source for the vorticity -- 
if this term is zero then in the absence of initial vorticity none will be sourced in the system. Thus,
for a barotropic fluid there is no source of vorticity. However, allowing for entropy in the system,
where the equation of state is then a function of two independent variables such as $P\equiv P(\rho,S)$,
provides a source of vorticity. This was first discovered by Crocco in Ref.~\cite{crocco}. In the remainder
of this article we extend this analysis to cosmology enabling us to study the sourcing of vorticity in the early
universe.

\section{Vorticity in Cosmology}

In order to study vorticity in the early universe we need to use general relativity. The standard, powerful
technique for modelling inhomogeneities in cosmology is to use cosmological perturbation theory. Starting
with a homogeneous and isotropic Friedmann-Lema\^itre-Robertson-Walker spacetime as a background,
small, inhomogeneous perturbations are added on top, which are then expanded in a series.\footnote{There
are many reviews on this topic, e.g., Refs.~\cite{ks, mfb, MW2008}. We cannot go into details in
this brief article.}
So, for example, the energy density is perturbed as
\begin{equation}
\rho(\vec{x},\eta)=\rho_0(\eta)+\delta\rho_1(\vec{x},\eta)+\frac{1}{2}\delta\rho_2(\vec{x},\eta)+\cdots\,,
\end{equation}
where the subscripts denote the order of the perturbation, and $\eta$ denotes conformal time. We
wish to consider perturbations in a system containing a perfect fluid with non-barotropic equation of state 
$P\equiv P(\rho,S)$, in which case the pressure perturbation can be expanded (to first order) as
\begin{equation}
\delta P_1=\frac{\partial P}{\partial S}\delta S_1+\frac{\partial P}{\partial \rho}\delta\rho_1\,.
\end{equation}
On introducing the non-adiabatic pressure perturbation, $\delta P_{{\rm nad}1}$, and the adiabatic sound speed
$c_{\rm s}^2$, this can be rewritten as $\delta P_1=\delta P_{{\rm nad}1}+c_{\rm s}^2\delta\rho_1$. This can
be extended beyond linear order (for details see, e.g., Ref.~\cite{nonad}). The line element for scalar and 
vector perturbations in the uniform curvature gauge takes the form
\begin{equation}
ds^2=a^2(\eta)\Big[-(1+2\phi)d\eta^2+2B_{,i}dx^id\eta+\delta_{ij}dx^idx^j\Big]\,.
\end{equation}

In general relativity we define the vorticity tensor as 
$\omega_{\mu\nu}={\cal P}_\mu{}^\alpha{\cal P}_\nu{}^\beta u_{[\alpha;\beta]}$ where ${\cal P}_{\mu\nu}$
is the projection tensor into the fluid rest frame, and $u^\mu$ is the fluid four-velocity. This can be 
expanded order-by-order in perturbation theory. On doing so, and using the evolution equations from 
energy-momentum conservation and the Einstein field equations, we can obtain the evolution equation for
first order vorticity,
$
\omega_{1ij}{}'-3{\cal H} c_{\rm s}^2\omega_{1ij}=0\,.
$
This reproduces the well known result \cite{ks} that, in a radiation dominated universe,
$|\omega_{1ij}\omega_1^{ij}|\propto a^{-2}$, in the absence of anisotropic stress. However, at second order
things are not as straightforward and we find, after a fair amount of algebra, that the second order
vorticity evolves as \cite{vorticity}
\begin{equation}
\omega_{2ij}{}'-3{\cal H}c_{\rm s}^2\omega_{2ij}=
\frac{2a}{\rho_0+P_0}\Big\{3{\cal H}V_{1[i}\delta P_{{\rm nad}1,j]}
+\frac{\delta\rho_{1,[j}\delta P_{{\rm nad}1,i]}}{\rho_0+P_0}\Big\}\,,
\end{equation}
in the absence of linear vorticity. 
Thus, we see that while a barotropic fluid does not source any vorticity,
on allowing for a non-zero adiabatic pressure, or isocurvature, perturbation, vorticity is sourced. This is a 
generalisation of Crocco's theorem to an expanding background. 

\section{Discussion}

We have briefly shown that vorticity can be sourced at second order in perturbation theory by a coupling
between scalar perturbations. This is one example of the importance of studying higher-order
perturbation theory in light of the forthcoming influx of experimental data, and shows the potential 
for  new phenomena in the higher-order theory giving rise to new observational predictions. The next goal
is to study the effect of this vorticity on the Cosmic Microwave Background (CMB).

In Ref.~\cite{Christopherson:2010ek} we made a first attempt at studying the importance of this vorticity
and found a power spectrum with large amplification on small scales. In order to perform this study
we made an ansatz for the non-adiabatic pressure perturbation. Future work will attempt to use
a more realistic input, for example from the isocurvature perturbations in multi-field
inflationary systems (e.g. Ref.~\cite{Huston:2011fr}), or from the relative entropy between
relativistic and non-relativistic species in the standard concordance cosmology (e.g. Ref.~\cite{Brown:2011dn})
in order to investigate the importance of this induced vorticity, and the possibility for this vorticity
to source B-mode polarisation of the CMB. Furthermore, since vorticity is intimately 
related to magnetic fields \cite{biermann}, another interesting possibility is the sourcing of 
primordial magnetic fields from second order vorticity \cite{nalson}.

In summary, as our ability to build powerful experiments to collect higher quality data sets improves,
we are in a position to study higher-order effects of cosmological perturbation theory, such as vorticity,
and investigate their importance to the physics of the early universe.

\section*{Acknowledgements}
The authors are grateful to David Matravers for an enjoyable collaboration on which this article is based. 
AJC is supported by the Sir Norman Lockyer Fellowship of the Royal Astronomical Society and acknowledges
the School of Physics and Astronomy and a Research Staff Travel Prize from the University of Nottingham
and an IUPAP grant through the MG13 organisers, for support to attend the conference. KAM is supported,
in part, by STFC grant ST/J001546/1.

\bibliographystyle{ws-procs975x65}
\bibliography{christopherson}

\begin{thebibliography}{10}

\bibitem{crocco}
L.~Crocco, {\em Z. Angew. Math. Mech} {\bf 17}, p.~1 (1937).

\bibitem{ks}
H.~Kodama and M.~Sasaki, {\em Prog. Theor. Phys. Suppl.} {\bf 78}, 1 (1984).

\bibitem{mfb}
V.~F. Mukhanov, H.~A. Feldman and R.~H. Brandenberger, {\em Phys. Rept.} {\bf
  215}, 203 (1992).

\bibitem{MW2008}
K.~A. Malik and D.~Wands, {\em Phys. Rept.} {\bf 475}, 1 (2009).

\bibitem{nonad}
A.~J. Christopherson and K.~A. Malik, {\em Phys. Lett.} {\bf B675}, 159 (2009).

\bibitem{vorticity}
A.~J. Christopherson, K.~A. Malik and D.~R. Matravers, {\em Phys. Rev.} {\bf
  D79}, p. 123523 (2009).

\bibitem{Christopherson:2010ek}
A.~J. Christopherson, K.~A. Malik and D.~R. Matravers, {\em Phys.Rev.} {\bf
  D83}, p. 123512 (2011).

\bibitem{Huston:2011fr}
I.~Huston and A.~J. Christopherson, {\em Phys.Rev.} {\bf D85}, p. 063507
  (2012).

\bibitem{Brown:2011dn}
I.~A. Brown, A.~J. Christopherson and K.~A. Malik, {\em
  Mon.Not.Roy.Astron.Soc.} {\bf 423}, p. 1411 (2012).

\bibitem{biermann}
L.~Biermann, {\em Z. Naturforsch. Teil {\bf{A}}} {\bf 5}, p.~65 (1950).

\bibitem{nalson}
E.~Nalson, A.~J. Christopherson and K.~A. Malik, {\em in preparation}  (2012).

\end{thebibliography}

\end{document}